\documentstyle[12pt]{article}

\begin{document}
\input epsf.tex

\begin{titlepage}
\begin{flushright}
IC/2000/62\\
hep-th/0005230\\
\end{flushright}

\begin{center}

{\Large\bf A Self-tuning  Exact Solution \\[5mm]
and the Non-existence of \\[5mm]
        Horizons in 5d  Gravity-Scalar System} \\[10mm]
{{\bf Chuan-Jie Zhu}\footnote{e-mail: zhucj@itp.ac.cn}\\[5mm]
    {\it Institute of Theoretical Physics, Chinese Academy of}\\
{\it Sciences, P. O. Box 2735, Beijing 100080,
P. R. China\footnote{Permanent address}}\\[3mm]
and } \\[3mm]
{\it Abdus Salam International Center for Theoretical Physics\\
Strada Costiera 11, I-34014 Trieste, Italy}\\[1cm]
{\bf ABSTRACT}\\[5mm]

\parbox{4.5in}{We present an exact thick domain wall solution with
naked singularities to five dimensional gravity coupled with a scalar
field with exponential potential. In our solution we found exactly the special
coefficient of the exponent as coming from compactification of string theory
with cosmological constant. We show that this solution is self-tuning when a
3-brane is included. In searching for a solution with horizon we found a
similar exact solution with fine-tuned exponent coefficient with an integration
constant. Failing to find a solution with horizon we prove the non-existence of
horizons. These naked singularities actually can't be resolved by horizon. 
We also comment on the physical relevance of this solution.}
\end{center}

\end{titlepage}

\section{Introduction}
 
After the Randall-Sundurm (RS) \cite{RSa,RSb} suggestion of a new
compactification to confine gravity to four-dimensions, there is an explosive
activity of studying various generalizations. One possible application of
the RS scenario is to solve  the cosmological constant
problem which is the key obstacle to make the models of
particle physics that can be derived from string theory more realistic
\cite{Weinberg,Witten}.
Recently a simple self-tuning mechanism has been
suggested in \cite{Kachrua, Kachrub, Arkani} to at least improve on the 
cosmological constant problem (see also
\cite{Rubakov} for an earlier mechanism involving
extra dimensions) where all order standard model loop contributions
are off-set by the parameters appearing in the solution to the five dimensional
gravity-scalar system. The authors of \cite{Kachrua, Kachrub, Arkani}
showed that one can find static
solutions to the classical equations of motion for the coupled gravity-scalar
system. However, all the solutions found in \cite{Kachrua, Kachrub, Arkani}
(see also \cite{Youm})  were  obtained either by having a constant
potential for the scalar or by making a simple ansatz.
In \cite{Csakia} some  integrable bulk potentials are obtained and 
a general and exact solution
was obtained for the exponential potential by using a first order formalism
\cite{Gubserb,Townsend, Gibbons, Giradellob,DeWolfe, 
Gremm,Csakib,Gubsera,Cvetic}.
Most of these solutions have naked 
singularities.

Leaving  aside the physical interpretation of the naked singularity one would 
like to understand
better these solutions. However the ``explicit'' solution given in \cite{Csakia}
is not quite explicit and it is given only 
as an implicit function of the coordinate $r$ (see below). Even if this is quite
enough for numerical analysis they are not quite suitable for ananlytic
calculations.
Some solutions \cite{Townsend, Gremm} are quite explicit and
reasonably simple.
But these solutions are just  invented purposely (to be solvable and simple)
and the origin for the scalar potential is not clear. (The potentials 
in gauged supergravity are also quite complicated  and the brane world is
not yet realized as a BPS or non-BPS configuration of supersymmetric theory 
\cite{Susya,Susyb,Susyc} (however see \cite{Behrndt}
for a cosine potential).) In this paper we will fill this gap and show that
there does exist a simple and exact 
solution with the simple exponential scalar potential. Amazingly, the
exponent for which we can find such a solution is exactly the value
which comes from string theory as noted in \cite{Kachrua}.

The organization of this paper is as follows: in Section 2 we present full
details of our exact
solution. Here the solution is not obtained by using the first order formalism 
\cite{Gubserb,Townsend, Gibbons, Giradellob,DeWolfe, Gremm,Csakib,
Gubsera,Cvetic} as was done in
\cite{Csakia}. Instead we cast the coupled system of equations into a
simple form of (effectively) two first order differential equations, given as 
eqs.~(\ref{eqsimpleaa}) and (\ref{eqsimpleba}).
From here we found the special value which made the equations solvable 
by elementary method and function. This value $c=1$ 
($a = \sqrt{2 \over 3}$) corresponds exactly to  the potential
which was reduced from the compactification of string theory with a
non-vanishing cosmological
constant \cite{Kachrua} ( $a = {4\over 3}$ in their 
normalization, see eq.~(\ref{eqzeroone}) below).
We also show that this solution is self-tuning \cite{Kachrua}
when a 3-brane is included. 
In Section 3 we turn to a more general problem of finding solutions
with horizons. Here the
equations can be simplified as in Section 2 and we also found a 
first integration, eq.~(\ref{crucial}). The
equations obtained have the same structure as the equations without horizons.
By using the same method we found
a similar exact solution with fine-tuned exponent coefficient with an
integration constant (eq.~(\ref{citehere})). Failing to find a solution 
with horizon  we prove the non-existence of
horizons. The naked singularities cannot be resolved by horizons. 
We comment and conclude in Section 4.

\section{The self-tuning exact solution}

Our starting point  is the following action for five-dimensional gravity
coupled to a single real scalar with an exponential potential \cite{Kachrua}:
\begin{equation}
S = \int d^D x\, \sqrt{|G|}\left(  R -
\frac{1}{2}(\nabla\phi)^2 - \Lambda \, e^{ a \phi} \right).
\label{eqzeroone}
\end{equation}
Brane sources are not included at this moment but they can easily be added
(see below).  From this action  we have the following equations of motion:
\begin{eqnarray}
& & R_{MN} -{1\over 2} \, 
\nabla_M \phi \, \nabla_N \phi =  {\Lambda \over D-2 } \, e^{ a \phi}
\, G_{MN},  
\label{eqone} \\
& & \nabla^2\phi - a \, \Lambda \, e^{ a \phi}  =0.
\label{eqtwo}
\end{eqnarray}

In this paper we will first study solutions with Poincar\'e-invariant
$(D-1)$-dimensional
spacetime and we have the following ansatz for the metric:
\begin{equation}
ds^2 = e^{2A(r)}\left( \eta_{\mu\nu}d x^{\mu} d x^{\nu} \right) + dr^2,
\end{equation}
where the function $A$ depends only on $r$\footnote{We use the mostly
positive convention for the metric.}.

Substituting this ansatz into eqs.~(\ref{eqone})-(\ref{eqtwo}), these
equations are given as follows (setting $D=5$):
\begin{eqnarray}
& & A'' + 4 (A')^2 + { \Lambda \over 3} \, e^{ a \phi}  =0, 
\label{eqthree} \\
& & 4 A'' + 4 (A')^2 + {1\over 2}\, (\phi')^2+ { \Lambda \over 3} 
\, e^{ a \phi}  =0, 
\label{eqfour} \\
& & \phi'' + 4 A' \phi' -{ a \, \Lambda} \, e^{ a \phi} =0.
\label{eqfive}
\end{eqnarray}
where prime denotes differentiation with respect to $r$,
and we have assumed that the scalar field also depends only on $r$.

By simple algebra the above three equations can be recasted into the
following form
\begin{eqnarray}
& & A'' + { 1\over 6} \, (\phi')^2 = 0, 
\label{eqsimplea} \\
& & (A'' + {1\over 3 a} \phi'') + 4 (A' + {1\over 3 a} \phi') \, A' = 0, 
\label{eqsimpleb} \\
& & A'' + 4 (A')^2 + { \Lambda\over 3}  \,  e^{ a \phi}  =0, 
\label{eqsimplec}
\end{eqnarray}
By doing some simple rescaling   and setting  $\phi = \sqrt{6 c } \, 
\Phi$ and  $a = \sqrt{2 c/3}$, the
above equations are simplified to the following form:
\begin{eqnarray}
& & A'' + c \, (\Phi')^2 = 0, 
\label{eqsimpleaa} \\
& & (A'' +  \Phi'') + 4 (A' +  \Phi') \, A' = 0, 
\label{eqsimpleba} \\
& & A'' + 4 (A')^2 + { \Lambda \over 3} \,  e^{ 2 c \Phi}  =0, 
\label{eqsimpleca}
\end{eqnarray}
Adding   together (\ref{eqsimpleaa}) and (\ref{eqsimpleba}) gives
the following equation
\begin{equation}
(  2 A'' + \Phi'') + (2 A' + \Phi')^2 + (c-1) (\Phi')^2 = 0.
\end{equation}
Now we observe that for $c=1$ the above equation simplifies and
we can solve it to obtain
\begin{equation}
2 A' + \Phi' = { 1 \over r -r_0 }, 
\end{equation}
where $r_0$ is an arbitrary (integration) constant.

Now we use the above relation in eq.~(\ref{eqsimpleaa}) to
eliminate $A$ to get  the following equation for $\Phi$ (remember $c=1$):
\begin{equation}
\Phi'' - 2 \, (\Phi')^2 + { 1 \over (r - r_0)^2} = 0.
\end{equation}
To solve this equation we first find a special solution by trying the 
ansatz $\Phi' = a_1/(r - r_0)$ to get $a_1 =-1$. Setting 
\begin{equation}
\Phi' = \tilde{\Phi}' - { 1 \over r - r_0},
\label{backphi}
\end{equation}
the equation for $\tilde{\Phi}$ is given as follows
\begin{equation}
\tilde{\Phi}'' + {4 \over r - r_0} \, \tilde{\Phi}' - 2 \, (\tilde{\Phi}')^2 = 0,
\end{equation}
which can easily be solved, first by dividing both sides by $\Phi'$ and
then combining the first two terms into a total derivative to yield 
\begin{equation}
{1 \over \tilde{\Phi}' (r - r_0)^4} = {2/3 \over (r-r_0)^3 } + \hbox{const.}.
\end{equation}
Substituting the above result back to eq. (\ref{backphi}) we finally get
\begin{eqnarray}
\Phi(r) & = & {1\over 2} \left( \ln|r-r_0| - \ln|(r_1-r_0)^3 - (r-r_0)^3|\right) 
+ \Phi_0,
\label{lastphi} \\
A(r) & = & {1\over 4} \left(\ln|r-r_0| + \ln|(r_1-r_0)^3 - (r-r_0)^3|\right)
+ A_0,
\label{lastA} 
\end{eqnarray}
where $r_0$, $r_1$, $\Phi_0$ and $A_0$ are integration constants.
Substituting these results back into eq. (\ref{eqsimpleca}) fixes
the integration constant $\Phi_0$:
\begin{equation}
\Phi_0 = {1\over 2 } \ln {9\over |\Lambda|} = \ln 3 - {1\over 2}\, 
\ln |\Lambda|.
\label{eqphizero}
\end{equation}
 
For $\Lambda>0$ the consistency of the equation ( $A'' + 4 (A')^2 \le 0$)
requires $0<r-r_0<r_1 - r_0$ for $r_1>r_0$ and
$r_1-r_0<r-r_0<0$ for $r_1<r_0$. It is always possible to set $r_0=0$ by
using the translation  invariance along the fifth dimension. We also assume
$r_1>0$ as the other case can be obtained
from this one by a reflection operation $r \to - r$.  In this case our solution
is interpolating between two naked singularities at $r=0$ and $r=r_1$.

On the other hand for $\Lambda<0$ the consistency of the equation
requires the following  inequality (after setting $r_0=0$):
\begin{equation}
{ r \over r^3  -r_1^3}   > 0. 
\end{equation}
For $r_1>0$ this gives two regions: $r>r_1$ and $r<0$. For $r_1<0$
the two regions are $r>0$ and  $r<r_1$. In both cases, the five dimensional
spacetime consists of two disconnected pieces. Each piece has a naked
singularity at $r=\pm r_0$ and extends to $\pm \infty$. 

\begin{figure}[ht]
    \epsfxsize=120mm%
    \hfill\epsfbox{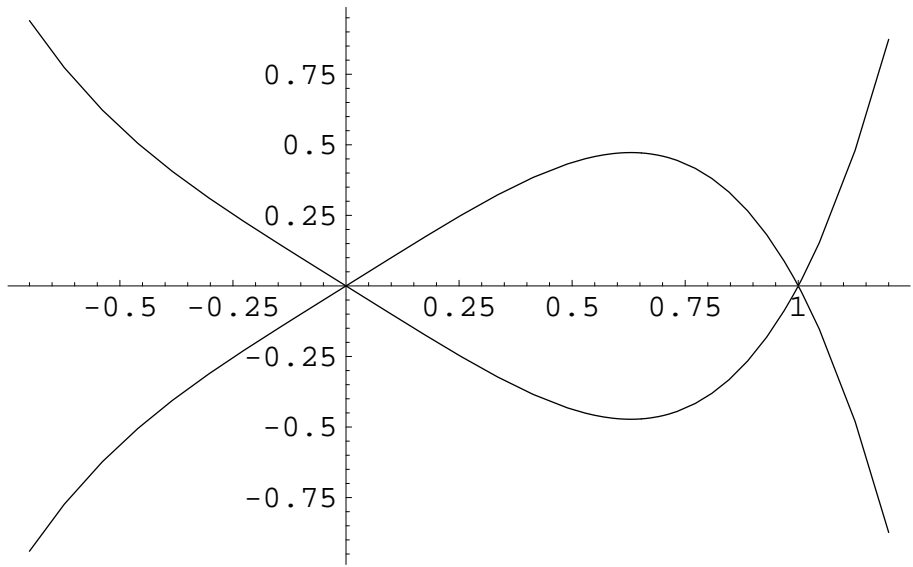}\hfill~\\
    \caption{The profile of 4d spacetime volume along the 5th dimension.}
    \label{figa}
   \end{figure}

To get some idea of how the solution looks like, in  Fig.~\ref{figa} we
show the profile of the  volume of the 4d spacetime   along the fifth
dimension $r$. The height is the relative volume of the 4d space-time which
is given by $e^{4 A(r)} = |r(1-r^3)|$. The middle part is the region
$\Lambda >0$. 
The disconnected left and right parts are the two regions for $\Lambda <0$.   

As one can see from Fig.~\ref{figa}, this solution is not symmetric
between the two naked singularities. The singularity ar $r=0$ is at 
weak coupling $g=e^{\phi}=0$, and the singularity at 
$r=R$ is at strong coupling $g = + \infty$. We will discuss the 
possible physical interpretation of this asymmetry in Section 4.

With this exact (bulk) solution at hand one can also study the self-tuning
mechanism of  refs. \cite{Kachrua,Kachrub,Arkani} by putting  a 3-brane 
along the 5th dimension, say at $r=0$. As the self-tuning mechanism is
not evident here,  we will analyse it  in some details.
For definiteness we will consider the case of $\Lambda<0$. The solution
on the two sides of the 3-brane is given as follows\footnote{This solution 
doesn't confine gravity because $A(r) \to + \infty$ as $r \to \pm \infty$. We
use it here just to  show the  self-tuning mechanism.}:
\begin{eqnarray}
r&>&0,  
\nonumber \\
& & A(r) = {1\over 4}\left(  \ln(r + r_1) + \ln( (r+r_1)^3 -R_1^3) \right)
+ A_1,
\\
& & \Phi(r) = {1\over 2} \left(  \ln(r + r_1) 
- \ln( (r+r_1)^3 -R_1^3) \right) + \Phi_0,
\\
 r&<&0,
\nonumber \\
& & A(r) = {1\over 4}\left(  \ln(r_2 - r) + \ln( (r_2-r)^3 -R_2^3) \right) 
+ A_2,
\\
& & \Phi(r) = {1\over 2} \left(  \ln(r_2 - r) 
- \ln( (r_2-r)^3 -R_2^3) \right) + \Phi_0,
\end{eqnarray}
where $r_1$, $r_2$, $R_1$, $R_2$, $A_1$ and $A_2$ are
constants and $\Phi_0$ is given by eq.~(\ref{eqphizero}).  We take $r_1$
and $r_2$ to be positive and they satisfy the following
inequalities: $r_1>R_1$ and $r_2>R_2$. ($R_1$ and $R_2$
are not necessarily positive and can take any (real) values.) One can check
that these are solutions to the bulk equations in their respective valid
regions.

The matching conditions are as follows \cite{Kachrua}\footnote{Please do the 
rescaling $\phi \to {3\over 2} \Phi$ and set $a = {4\over 3}$.}
\begin{eqnarray}
& & \Phi'(r)|_{r=0+} - \Phi'(r)|_{r=0-} = { b \, V \over 4}\, e^{ {3 b\over 2} 
\Phi(0)},
\\
& & A'(r)|_{r=0+} - A'(r)|_{r=0-} = -{  V \over 6 }\, e^{ {3 b\over 2} 
\Phi(0)},
\\
& & 
\Phi(r)|_{r=0+} = \Phi(r)|_{r=0-} \equiv \Phi(0),
\\
& & A(r)|_{r=0+} = A'(r)|_{r=0-} .
\end{eqnarray}
One can always use the last equation to solve $A_1$ and the other constant
$A_2$ is an overall  constant. Noticing the particular
combinations $2 A' \pm  \Phi'$ we obtain the following relations
from the first three equations in the above:
\begin{eqnarray}
& & { 1\over r_1} + { 1\over r_2} = - { V\over 4} \, ( { 4 \over 3} - b) 
\, e^{ {3 b\over 2} \Phi(0)},
\label{lastzz}
\\
& & { 3 \, r_1^2 \over r_1^3 - R_1^3 } 
+ { 3 \, r_2^3 \over r_2^3 - R_2^3} 
= - { V\over 4} \, ( { 4 \over 3} + b) \, e^{ {3 b\over 2} \Phi(0)},
\label{lastzx}
\\
& & \Phi(0) - \Phi_0 = 
{ 1\over 2} ( \ln r_1  - \ln( r_1^3 - R_1^3) ) 
= { 1\over 2} ( \ln r_2  - \ln( r_2^3 - R_2^3) ).
\label{lastzy}
\end{eqnarray}
Because of $r_{1,2}>0$, we must have $V<0$ and
$-{4\over 3}<b<{4\over 3}$ in order to have solutions. 
From eq. (\ref{lastzy}) one can solve $R_2$ in terms of $r_{1,2}$
and $R_1$ uniquely. By using eq.~(\ref{lastzy}), 
Eq. (\ref{lastzx}) can be simplified as follows:
\begin{equation}
{ 3 \, r_1 \over  r_1^3 - R_1^3 }\, (r_1 + r_2 ) = 
- { V\over 4} \, ( { 4 \over 3} + b) \, e^{ {3 b\over 2} \Phi(0)},
\end{equation}
which can be combined with eq.~(\ref{lastzz}) to yield
\begin{equation}
r_2 = {(4 + 3 b) \over 3(4 - 3 b)} \, { r_1^3 - R_1^3 \over r_1^2} >0.
\end{equation}
Substituting this result into eq.~(\ref{lastzz}) and using eq.~(\ref{lastzy})
we have
\begin{equation}
{ 3 \over r_1^{ ( 1 + { 3 b \over 4})} (r_1^3 
- R_1^3)^{ ( 1 - { 3 b \over 4})} }\,
\left[ \left( 1 +  {(4 + 3 b) \over 3(4 - 3 b)} \right) \, r_1^3 
- R_1^3 \right] = 
- { V \over 4 } \, ( {4 \over 3 } - b) \, e^{ { 3b \over 2 } \, \Phi_0} .
\end{equation}
For $ -{4\over 3} <b <{ 2 \over 3}$ and  taking any value of $R_1>0$,
as $r_1$ varies from $R_1$ to $+\infty$, the left-hand side of
the above equation varies from $+\infty$ to $0$. So for any value ($>0$) 
of the right-hand side, we
can find a value of $r_1>0$ which solves the above equation. This will give
a unique solution with $r_2$ and $R_2$ which depend on an arbitrary
$R_1$ and the various parameters appearing in the action: $V$, $b$ and
$\Lambda$.  So we have a flat 4 dimensional spacetime which
is self-tuning. Presumably other regions of the parameters will be covered by
different choices of pasting the bulk solutions \cite{Kachrua}.

As a final note, we point out that we can also
get the solution (II) obtained in \cite{Kachrua} by taking the limit
$\Lambda \to 0$. Here we must adjust the integration constant $r\equiv R$
appropriately with $\Lambda$ so that
one can take the limit smoothly. An expansion of the solution 
\begin{eqnarray}
\Phi(r) & = & {1\over 2} \left( \ln|r| - \ln|R^3 - r^3 |\right) + \Phi_0,
\label{lastlastphi} \\
A(r) & = & {1\over 4} \left(\ln|r| + \ln|R^3 - r^3|\right)+ A_0,
\label{lastlastA} 
\end{eqnarray}
around $r=0$ gives the solution (2.37) and (2.38) with the positive sign
in \cite{Kachrua},  and an expansion around $r=R$ gives the solution with
the minus sign there.

\section{The non-existence of solutions with horizon}

It is important to understand better the physical significance of the
singularities in our solution.
Normally these naked singularities would be discarded as unphysical,
but in some instances there are reasons to believe that considering these
singularities may be meaningful \cite{Gremm,Gubsera}. 

One possible resolution of these naked singularities is to cover them with
horizons which have been studied in the appendix of \cite{Gubsera}.
Here we will prove the non-existence of such solutions with (generic)
exponential potential. We will comment on other interpretations of
these singularities in Section 4.

As in the Schwarzchild solution in general relativity we make the following
ansatz for  the 5 dimensional metric \cite{Gubsera}:
\begin{eqnarray}
ds^2 & =  & - e^{2A_0(r)} (d t)^2 +  e^{ 2 A_1(r)} ( (d x^1)^2 + 
(d x^2)^2 + ( d x^3)^2 ) + e^{ 2(A_1(r) - A_0(r) } (d r)^2
\nonumber \\
&  = &e^{ 2 A_1(r)} \left( - h(r) (d t)^2 +  (d x^1)^2 + 
(d x^2)^2 + ( d x^3)^2  \right) + { (d r)^2 \over h(r) },
\end{eqnarray}
where $h(r) = e^{2A_0(r)- 2A_1(r)}$.
Horizon appears at the point where the function $h(r)$ has a simple zero.

By using this ansatz we obtain the following equations of motion from 
eqs. (\ref{eqone})-(\ref{eqtwo}):
\begin{eqnarray}
& & A_0'' + 2 \, A_0'\, (A_0' + A_1') + { \Lambda \over 3}
\, e^{ a \phi + 2 (A_1 - A_0)} = 0,
\\
& & A_1'' + 2 \, A_1'\, (A_0' + A_1') + { \Lambda \over 3}
\, e^{ a \phi + 2 (A_1 - A_0)} = 0,
\\
& & A_0'' + 3 A_1'' +2 \, A_0'(A_0' + A_1') + {1\over 2} \, (\phi')^2 + 
{ \Lambda \over 3} \, e^{ a \phi + 2 (A_1 - A_0)} = 0,
\\
& & \phi'' + 2 \, \phi'\, (A_0' + A_1') 
- {  a \, \Lambda } \, e^{ a \phi + 2 (A_1 - A_0)} = 0,
\end{eqnarray}
A similar rescaling of the field $\phi$ and some simple algebras
can bring the above equations to the following equivalent form:
\begin{eqnarray}
& & A_1'' + c \, (\Phi')^2 = 0, 
\label{heqsimpleaa} \\
& & (A_1'' +  \Phi'') + 2 (A_1' +  \Phi') \,( A_0'+ A_1') = 0, 
\label{heqsimpleba} \\
& & (A_0''-A_1'')  + 2 (A_0'-A_1')\,( A_0'+ A_1') = 0, 
\label{heq}\\
& & A_1''  + 2 \, A_1' (A_0' + A_1') + { \Lambda \over 3} 
\,  e^{ 2 c \Phi+ 2 (A_1- A_0)}  =0, 
\label{heqsimpleca}
\end{eqnarray}
Now we can use eq. (\ref{heqsimpleba}) and eq. (\ref{heq}) to get
\begin{equation}
A_0' - A_1' = d\, (A_1' + \Phi'), 
\label{crucial}
\end{equation}
where $d$ is an integration constant. By using this relation we can
eliminate $A_0'$ from eqs. (\ref{heqsimpleaa})-(\ref{heqsimpleba})
to arrive at a similar system of equations as discussed in the previous section. 
We will not  go through the various steps as we did in the previous section,
as we have intentionally presented the full details arriving at the special
value and solving the resulting equations there.
The final result is that if we choose $d$ carefully (a fine-tuning) we
can actually use the method in Section 2 to obtain an  exact solution. 
We have
\begin{eqnarray}
d  & = & { 2(c-1)\over 2 -c}, 
\label{citehere}
\\
\Phi (r) & = & { (2-c) \over 2 c} \, \ln |r | - {1\over 2} \ln \left| 
 r_1^{({4 \over c} -1)} -  r^{({4 \over c} -1)} \right| + \Phi_0, \\
A_0(r) & = & -{(c^2 - 6 c +4)\over 4 c} \, \ln |r| 
\nonumber \\
& & \qquad +  {(2-c)\over 4}\, 
\ln \left|  r_1^{({4 \over c} -1)} -  r^{({4 \over c} -1)} \right| +a_0,
\\
A_1(r) & = & { (2-c)^2\over 4 c}    \, \ln |r| + { c \over 4} 
\, \ln \left| r_1^{({4 \over c} -1)} -  r^{({4 \over c} -1)} \right| +a_1.
\end{eqnarray}
Note that the above solution is not valid for $c=2$, for this will give
$d = \infty$ which is not a  well defined mathematical operation. In this
special case we have  $A_1' = - \Phi'$. This kind of ansatz was used
in \cite{Kachrua} and has been well studied there. 

With the above special solution at hand one can easily see that there
is no horizon. The function
$h(r)$ is given as
\begin{equation}
h(r) \equiv e^{ 2 (A_0(r) - A_1(r))} = 
\hbox{(const.)}\times r^{ -(c^2 -  5c + 4)\over 2 c}\,  (r_1^{({4\over c} -1)}
 - r^{ ({ 4\over c} - 1)})^{ 1-c},
\end{equation}
which can  only have possible zeroes at $r=0$ and/or $r= r_1$.
These two points are also the singularities of the metric (curvature).
So it is not a horizon.

Actually one can prove that there is  no exact solutions with horizons. This
proof doesn't depend on the above special solution. The crucial observation
is eq. (\ref{crucial}). Assuming we do have a solution with a horizon at
$r=r_h$:
\begin{equation}
h(r) = e^{ 2 (A_0(r) - A_1(r))} = a_3( ( r - r_h) + O((r-r_h)^2) ).
\end{equation}
Then we have
\begin{equation}
A_0' - A_1' = { 1/2 \over r -r_h} + \cdots,
\end{equation}
around $r= r_h$. If we require that $r_h$ is different from any singularities
of $A_1(r)$, $A_1(r)$ is well behaved around $r=r_h$ and so is $A_1'$:
\begin{equation}
A_1'(r) = A_1'(r_h) + A_1''(r_h)(r - r_h) + \cdots.
\end{equation}
By using eq. (\ref{crucial}) we have the following expansion for $\Phi'$:
\begin{equation}
\Phi'(r) = { 1\over 2 d} \, { 1\over r -r_h} + \cdots, 
\end{equation}
but this expansion is in sharp contradiction with
eq.~(\ref{heqsimpleaa}) ($c\neq 0$).
So we proved that there is no solution with horizon  for the five dimensional
gravity coupled with a scalar with exponential
potential. For a discussion with a generic potential with more scalar fields,
please see \cite{Gubsera}.

\section{Comment and conclusion}

Five dimensional gravity (with or without matters) is an interesting subject
which was resurrected many times. The original Kaluza-Klein idea
involves only 5 dimensions. But non-abelian gauge field requires
more dimensions. With the advent of superstring and M theory, one can view
the extra dimensions as a dynamical thing: the radii of the extra dimensions
could depend on coupling constant. Recent development with AdS/CFT
correspondence \cite{Maldacenaand} gives a new 
look at the fifth dimension: it is identified with the energy scale of the 4d 
field theory  \cite{Wittenc,Polchinski}  and the Hamiltonian-Jocobi equation
was used to derive the renormalization group flow equations
\cite{Verlindea,Verlindeb} (see \cite{Giradelloa,Khavaev,Karch,Gubserb,
Gubserc} for related works and, for example,
\cite{Verlinded,Sahakian,Verlindee,Verlindef,Wu} for
later developments). In RS scenarios the compactification and the confining
of gravity to  4 dimensions was achieved by an exponential warped factor
and one has the freedom to put 3-branes at some points along the fifth
dimension by fine-tuning various parameters \cite{RSa,RSb}. According to
\cite{Kachrua,Kachrub,Arkani} this  fine-tuning is  not required
because the parameters in the solution will adjust themselves
(a self-tuning mechanism). But  here we have a plethora of different solutions
and some solutions have naked singularities. The puzzle is: what principle
should we use to select the right  solution?

As we said in the introduction, solutions with singularities cannot be simply
discarded as unphysical. Some solutions with singularities have physical
interpretations in type IIB superstring theory (see \cite{Gubserb,Polchinskib},
for example).
In fact the solution we found in eqs.~(\ref{lastphi})-(\ref{lastA}) was argued
as some solution-independent  behaviour \cite{Giradelloa} and corresponds
to the deformation of non-conformal field theory
\cite{Kehagias,Gubserc}. We may associate the two naked singularities to two
different field theories (or putting
it differently: the  two naked singularities are
resolved by two well-defined field theories).  The problem with this
interpretation is that it is difficult to think that these two non-conformal field
theories can be related by a renormalization group flow. The different
behaviour of the coupling constants between the two singularities offer a
possible way out: either the 5d gravity description breaks down at one of the
singularities or these two non-conformal field theories
are in fact strong-weak dual pairs. It is interesting to study the higher
embedding of these naked singularities. 

For $\Lambda>0$ our exact solution has an interesting behaviour: as $r$
varies from $0$ to $r_1$, $A(r)$ (so is $e^{4 A(r)}$) first increases, 
reaches a maximum, and then decreases, all within a finite interval of $r$.
In the original RS scenario this behaviour was introduced by hand by 
introducing 3-brane. Here we have no 3-branes. We can only attribute
this behaviour to the inclusion of scalar field. In the AdS/CFT
correspondence \cite{Maldacenaand}, scalar fields are interpreted as
deformations of four dimensional field theory. The profile of these scalar
fields are just  running coupling constants. This correspondence is valid
in the strong coupling. Recent suggestions to extend this correspondence
to weak coupling seem to be in
contradiction with our exact solution. A general programme of
Hamiltonian-Jacobi equation/Renormalization group flow equation
correspondence seems only an approximation.

\section*{Acknowledgments}

I would like to thank Roberto Iengo, Shamit Kachru  and K. S. Narain
for helpful discussions and comments. This work is  supported in part by
funds from National Natural Science  Foundation of China and Pandeng
Project. The author would also like to thank Prof. S. Randjbar-Daemi  and
the hospitality at Abdus Salam International Centre for Theoretical Physics,
Trieste, Italy.

\end{document}